\documentclass[2column]{IEEEtran}

\usepackage{epsfig}

\usepackage{amsthm}
\usepackage[cmex10]{amsmath}
\usepackage{color}
\usepackage{graphicx}
\usepackage{citesort}
\usepackage{amsfonts}
\usepackage{amssymb}
\usepackage{bbm}
\usepackage{dsfont}
\usepackage{times}
\usepackage{latexsym}
\usepackage{graphicx}
\usepackage{amssymb}
\usepackage{amsfonts}
\usepackage{psfrag}
\usepackage{subfigure}
\usepackage{multicol}
\usepackage{algorithmic}
\usepackage{algorithm}

\theoremstyle{definition} 

\theoremstyle{remark} 
\theoremstyle{remark} 
\theoremstyle{remark} 

\newcommand{\st}{\textrm{s.t.}}

\newcommand{\eps}{\varepsilon}
\newcommand{\E}{\textsf{E}}

\renewcommand{\H}{\textsf{H}}

\newcommand{\tr}{\textrm{tr}}

\newcommand{\brc}[1]{\left( #1 \right)}

\newcommand{\openone}{\leavevmode\hbox{\small1\normalsize\kern-.33em1}}

\newcommand{\bx}{\boldsymbol{x}}

\newcommand{\bn}{\boldsymbol{n}}

\newcommand{\bH}{\boldsymbol{H}}

\newcommand{\bI}{\mathbf{I}}
\newcommand{\bP}{\boldsymbol{P}}

\newcommand{\bzero}{\boldsymbol{0}}

\def\calC{{\mathcal{C}}}
\newcommand{\calN}{\mathcal{N}}

\newcommand{\main}{\textrm{M}}
\newcommand{\eave}{\textrm{E}}
\newcommand{\Rs}{R_{\textrm{s}}}
\newcommand{\Cs}{C_{\textrm{s}}}

\newcommand{\bHmH}{\boldsymbol{H}_{\main}^{\H}}
\newcommand{\bHeH}{\boldsymbol{H}_{\eave}^{\H}}
\newcommand{\bHm}{\boldsymbol{H}_{\main}}
\newcommand{\bHe}{\boldsymbol{H}_{\eave}}

\newcommand{\bym}{\boldsymbol{y}_{\main}}
\newcommand{\bye}{\boldsymbol{y}_{\eave}}
\newcommand{\bnm}{\boldsymbol{n}_{\main}}
\newcommand{\bne}{\boldsymbol{n}_{\eave}}

\newcommand{\Nm}{N_{\main}}
\newcommand{\Ne}{N_{\eave}}
\newcommand{\rhom}{\rho_{\main}}
\newcommand{\rhoe}{\rho_{\eave}}

\renewcommand{\bx}{\boldsymbol{x}}
\newcommand{\bxH}{\boldsymbol{x}^{ \H}}

\newcommand{\complex}[1]{\mathds{C}^{#1}}

\begin{document}
\IEEEoverridecommandlockouts
\sloppy

\title{On the Optimal Precoding for MIMO Gaussian Wire-Tap Channels}
\author{
\authorblockN{ Arash Khabbazibasmenj$^{1}$, Maksym~A.~Girnyk$^{2}$,
Sergiy A. Vorobyov$^{1,3}$, Mikko Vehkaper\"{a}$^{2,3}$, Lars K. Rasmussen$^{2}$\\}
\authorblockA{$^{1}$University of Alberta, Department of Electrical and
Computer Engineering, Alberta, Canada \\
email: \{\begin{tt}khabbazi, svorobyo\end{tt}\}\begin{tt}@ualberta.ca\end{tt}\\
 $^{2}$KTH Royal Institute of
Technology, School of Electrical Engineering and the ACCESS Linnaeus
Center,
Sweden\\
email: \{\begin{tt}mgyr, lkra\end{tt}\}\begin{tt}@kth.se\end{tt}\\
$^{3}$Aalto University, School of Electrical Engineering,
Department of Signal Processing and Acoustics, Finland\\
email: \{\begin{tt}mikko.vehkapera, sergiy.vorobyov\end{tt}\}\begin{tt}@aalto.fi\end{tt}}
\thanks{This work was supported by the Swedish Research Council (VR) and the National Science and Engineering Research Council (NSERC) of Canada.}
}

\maketitle

\begin{abstract}
We consider the problem of finding secrecy rate of a
multiple-input multiple-output (MIMO) wire-tap channel. A
transmitter, a legitimate receiver, and an eavesdropper are
all equipped with multiple antennas. The channel states from
the transmitter to the legitimate user and to the eavesdropper
are assumed to be known at the transmitter. In this contribution,
we address the problem of finding the optimal precoder/transmit covariance
matrix maximizing the secrecy rate of the given wire-tap channel.
The problem formulation is shown to be equivalent to a difference
of convex functions programming problem and an efficient algorithm
for addressing this problem is developed.
\end{abstract}

\section{Introduction}\label{sec:intro}
\begin{figure}
\centering
  \psfrag{A}[][][1]{\textsf{Alice}}
  \psfrag{B}[][][1]{\textsf{Bob}}
  \psfrag{E}[][][1]{\textsf{Eve}}
  \psfrag{Hm}[][][1]{$\bH_{\main}$}
  \psfrag{He}[][][1]{$\bH_{\eave}$}
  \psfrag{nm}[][][1]{$\bn_{\main}$}
  \psfrag{ne}[][][1]{$\bn_{\eave}$}
\includegraphics[width=\linewidth]{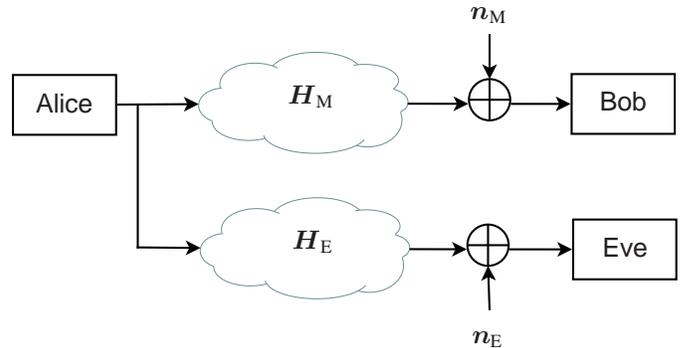}
  \caption{MIMO wire-tap channel.}
  \label{fig:wiretapChannel}
  \vspace{-0.5cm}
\end{figure}

In the recent years, the field of wireless physical layer security
has received considerable attention. Due to its inherent
randomness, wireless channels can be used for enhancing the secrecy
of communication. On the other hand, broadcast nature of a
wireless medium creates possibility for illegitimate parties to
eavesdrop the transmission. The corresponding basic setup (see
Fig.~\ref{fig:wiretapChannel}), referred to as the
\emph{wire-tap channel}, was firstly introduced by Wyner
in~\cite{wyner1975wire}. The \emph{secrecy rate}, introduced as a
performance metric for this setup, reflects the amount
of information per channel use that a source can reliably
transmit to a destination, provided that an eavesdropper does
not get any information. To achieve strictly positive secrecy rates, legitimate parties
must have statistically better channel than that of the eavesdroppers'.
Therefore, many techniques have been proposed to enhance the performance
of secure communication systems, e.g., multiple-input multiple-output
(MIMO) communications~\cite{shafiee2009towards},~\cite{liu2009note}. As
an extension, a scenario with multiple eavesdroppers has been studied
in~\cite{khisti2010misome},~\cite{oggier2011secrecy}.

The \emph{secrecy capacity} is defined as the maximum achievable secrecy
rate, and its evaluation is, in general, problematic due to the
non-convex nature of the corresponding optimization problem.
In the context of MIMO communications, the attempts of finding the
secrecy-capacity achieving precoding/covariance matrices have been
made already in~\cite{khisti2010misome}, \cite{oggier2011secrecy}.
Later, in~\cite{li2010optimal}, the secrecy capacity has been
characterized for the case of channel matrices with
certain rank properties. The special case with two-antenna
legitimate parties and a single-antenna
eavesdropper has been addressed
in~\cite{shafiee2009towards}.

Recently, several computationally efficient approaches have been proposed for
tackling the problem. For instance, in~\cite{fakoorian2012optimal},
a beamforming method based on the \emph{generalized
singular value decomposition} (GSVD) was proposed. Later, in~\cite{wang2012slnr},
two other sub-optimal approaches have been studied. A \emph{zero-forcing} (ZF)
precoder nulls out the leakage to the eavesdropper, thereby increasing the
secrecy rate, whereas the other approach maximizes the
\emph{signal-to-leakage-plus-noise} ratio (SLNR). The three approaches outperform
each other in various particular settings. Nevertheless, there yet exists
no unified approach allowing to compute the secrecy capacity achieving
transmit covariance matrix for the general MIMO setting.

In this paper, we study the non-convex problem of secrecy rate
maximization under the total power constraint in its most general
formulation. A widely used assumption
that the channel states from the transmitter to both the legitimate user and
eavesdropper are known at the transmitter~\cite{khisti2010misome},~\cite{fakoorian2012optimal}
is adopted here and the corresponding problem is formulated
as a difference of convex functions \emph{(DC) programming} problem.
We then develop an efficient algorithm for addressing the problem.
The proposed algorithm is based on the eigenvalue decomposition of
the transmit covariance matrix and a subsequent alternate optimization of
the eigenvectors and eigenvalues of the matrix. The
numerical results show that the proposed method outperforms the other
existing alternatives.

\section{System Model and Problem Formulation}\label{sec:sysMod}

The scenario of interest, depicted in Fig.~\ref{fig:wiretapChannel},
consists of the following two channels: the \emph{main channel}
between transmitter (Alice) and legitimate receiver (Bob), and
\emph{eavesdropper's channel} between Alice and the eavesdropper
(Eve). The corresponding input-output relations are given by
\begin{subequations}
\begin{align}
\bym =&\; \bHm \bx + \bnm, \\
\bye =&\; \bHe \bx + \bne,
\end{align}
\end{subequations}
where $\bx\sim \calC\calN(\bzero_{M},\bP)$ is the channel input
vector, $\bnm \sim \calC\calN(\bzero_{\Nm},\bI_{\Nm})$ and $\bne
\sim \calC\calN(\bzero_{\Ne},\bI_{\Ne})$ are additive noise
vectors at the receivers of the legitimate user and eavesdropper,
respectively. Here $\bI$ denotes the identity
matrix of the size given by its subscript. The entries of the channel matrices
$\bHm \in \complex{\Nm \times M}$ and $\bHe \in \complex{\Ne \times M}$
are assumed to be independent identically distributed according to
$\calC\calN(0,\rhom/M)$ and $\calC\calN(0,\rhoe/M)$, respectively,
where $\rhom$ and $\rhoe$ represent the corresponding
signal-to-noise ratios (SNRs). The channel state information (CSI),
consisting of the two above matrices, is assumed to be
perfectly known at the transmitter. Furthermore, the total
power constraint $\E \{ \tr\{\bx \bxH\} \} \leq M$
is employed at the transmitter. Here $\E\{\cdot\}$, $\tr\{\cdot\}$ and
$(\cdot)^{\H}$ stand for the mathematical expectation, trace of a matrix
and Hermitian transpose, respectively.

Based on the available CSI, it is possible to determine
the optimal covariance matrix $\bP$ that
maximizes the achievable secrecy rate, formally given as follows.
Let $w$ be a confidential message with entropy $H(w)$, which Alice wants to
communicate to Bob, and let $p_{e,n}$ be the probability of error
at Bob's receiver. Then a (weak) secrecy rate $\Rs$ is achievable
if there exists a sequence of $(2^{n \Rs},n)$ codes, such that
$p_{e,n} \to 0$ and $\frac{1}{n}H(w|\bye) \leq \frac{1}{n}H(w) -
\eps_n$, for some $\eps_n$ that tends to zero as $n\to\infty$.

Given CSI $\{\bHm,\bHe\}$ and transmit covariance matrix $\bP$, the achievable secrecy rate is given by~\cite{liu2009note}
\begin{subequations}
\begin{align}
 \Rs =&\; I(\bx; \bym|\bHm) - I(\bx; \bye|\bHe)\\
 =&\;\ln \det \brc{\bI_{\Nm} +
\bHm \bP \bHmH}  \nonumber\\
&\qquad\qquad\qquad\quad - \ln \det \brc{\bI_{\Ne} + \bHe \bP
\bHeH}.
\end{align}
\end{subequations}
The corresponding secrecy capacity is then obtained by
\begin{equation}
\Cs = \underset{\substack{\bP: \tr\{\bP\}\leq M \\
\bP \succeq \bzero_M}}{\max} \Rs, \label{problem}
\end{equation}
where $\bP \succeq \bzero_M$ means that $\bP$ is
a positive semi-definite matrix.

Since the objective function of the above problem is
\emph{not concave}, the problem cannot be solved directly with aid of
convex optimization tools. The GSVD approach is known to be efficient
for addressing the class of DC programming problems that~\eqref{problem}
belongs to \cite{fakoorian2012optimal}. In
\cite{khabbazibasmenj2012sum}, however, we have developed a more
general approach, which allows for deriving analytic results about
the global optimality. The corresponding algorithm is able to solve a
class of DC programming problems in polynomial time and is therefore
referred to as \emph{POlynomial Time DC} (POTDC) method. In
\cite{khabbazibasmenj2012sum}, the algorithm is applied to the problem
of optimal amplification matrix design for a two-way amplify-and-forward
network. The terminals therein are equipped with a single antenna and
only the relay has multiple antennas, while in~\eqref{problem} all
terminals are equipped with multiple antennas, thereby making the problem
more difficult. In what follows, we develop a novel
approach for addressing problems of type~\eqref{problem}.

\section{Main Result}\label{sec:mainResult}
Let $\boldsymbol P = \boldsymbol U^\textsf{H} \boldsymbol \Lambda
\boldsymbol U$ be the eigenvalue decomposition of the covariance
matrix $\boldsymbol P$, where $\boldsymbol U_{M \times M}$ is a
unitary matrix and $\boldsymbol \Lambda_{M \times M} \triangleq {\rm diag}
(\lambda_1, \lambda_2, \cdots, \lambda_M)$ is a diagonal matrix
whose elements are non-negative and correspond to the
eigenvalues of the matrix $\boldsymbol P$. Substituting $\boldsymbol
P = \boldsymbol U^\textsf{H} \boldsymbol \Lambda \boldsymbol U$ into
\eqref{problem} results in the following optimization problem
\begin{eqnarray}
\!\!\!\!\!\!\!\!&\max\limits_{\boldsymbol U, \boldsymbol \Lambda}&\!\!\!
\ln \det (\bI_{\Nm} + \bHm \boldsymbol U^\textsf{H} \boldsymbol
\Lambda \boldsymbol U \bHmH ) \nonumber \\
&& \quad \quad \quad \quad - \ln \det (\bI_{\Ne} + \bHe \boldsymbol
U^\textsf{H} \boldsymbol \Lambda \boldsymbol U  \bHeH ) \nonumber \\
&\st& \! \!\!{\rm tr} \{ \boldsymbol \Lambda \}\! \leq\! M, \ \boldsymbol
U^\textsf{H} \boldsymbol U\! =\!\bI_M, \  \lambda_i\! \geq\! 0, \ i\!=\!1,
\ldots , M \IEEEeqnarraynumspace \label{newform1}.
\end{eqnarray}
Let us now use Sylvester's determinant identity,
which implies that for any arbitrary matrices $\boldsymbol
A_{M \times N}$ and $\boldsymbol B_{N \times M}$, the following equality
holds~\cite{akritas1996various}
\begin{equation}
{\rm det}(\bI_M + \boldsymbol A \boldsymbol B) = {\rm det}(\bI_N +
\boldsymbol B \boldsymbol A).
\end{equation}
Thus, \eqref{newform1} can be equivalently
written as
\begin{eqnarray}
&\max\limits_{\boldsymbol U, \boldsymbol \Lambda}& \!\!\!\!\ln \det
(\bI_{\Nm} \!\!+\!\! \bHm \boldsymbol U^\textsf{H} \boldsymbol \Lambda
\boldsymbol U \bHmH ) \!-\! \ln \det (\bI_{M} \!\!+\!\! \boldsymbol D
(\boldsymbol U) \boldsymbol \Lambda ) \nonumber \\
&\st&  {\rm tr} \{ \boldsymbol \Lambda \} \!\leq\! M, \ \boldsymbol
U^\textsf{H} \boldsymbol U \!=\! \boldsymbol I_M, \  \lambda_i \geq 0, \
i\!=\!1, \ldots, M \IEEEeqnarraynumspace  \label{newform1.5},
\end{eqnarray}
where $\boldsymbol D (\boldsymbol U) \triangleq \boldsymbol U \bHeH
\bHe \boldsymbol U^\textsf{H}$ is defined for notation simplicity.
By a careful inspection of problem \eqref{newform1.5}, one can
readily observe that it is a DC programming problem over
the matrix $\boldsymbol \Lambda$ for a fixed value of $\boldsymbol
U$. It can be addressed using the POTDC algorithm
\cite{khabbazibasmenj2012sum}. Moreover, for a fixed $\boldsymbol
\Lambda$, \eqref{newform1.5} is an optimization problem
over unitary matrices that has been comprehensively studied in the
literature (see, e.g.,~\cite{Koivunen}). Based on the latter fact,
\eqref{newform1.5} can be addressed via alternating optimization, by first optimizing
with respect to $\boldsymbol \Lambda$ for a fixed value of $\boldsymbol U$,
and then further
optimizing with respect to $\boldsymbol U$ with $\boldsymbol \Lambda$
being set to the optimal value from the previous iteration. These
alternations continue until convergence as it is explained later
in the paper.

Optimization problem \eqref{newform1.5}, when $\boldsymbol U$ is
fixed to $\boldsymbol{U}_0$, can be expressed as
\begin{eqnarray}
&\max\limits_{\boldsymbol \Lambda}& \!\!\!\!\ln \det (\bI_{\Nm} +
\bHm \boldsymbol U_0^\textsf{H} \boldsymbol \Lambda \boldsymbol
U_0 \bHmH ) \nonumber \\
&& \quad \quad \quad \quad - \ln \det (\bI_M + \boldsymbol D
(\boldsymbol U_0) \boldsymbol \Lambda ) \nonumber \\
&\st&  {\rm tr} \{ \boldsymbol \Lambda \} \leq M, \ \lambda_i
\geq 0, \ i=1, \ldots, M \label{newform2}.
\end{eqnarray}
The resulting problem \eqref{newform2} is addressed via the POTDC
algorithm, which is efficiently used for the DC programming
problems, whose non-concave part is a function of a single variable.
Since the non-concave part of problem \eqref{newform2} is
not a function of a single variable, we instead maximize a lower
bound on the objective function of~\eqref{newform2}, whose
non-concave parts depend on a single variable each. For this goal, we
utilize the Hadamard's inequality \cite{horn2012matrix}, which implies
that the determinant of a Hermitian positive semidefinite matrix is
upper-bounded by the product of its diagonal elements. By applying
this inequality, the optimization problem of maximizing the lower
bound of the objective function of~\eqref{newform2} can
be recast as
\begin{eqnarray}
&\max\limits_{\boldsymbol \Lambda}& \!\!\!\!\ln \det (\bI_{\Nm}
\!+\! \bHm \boldsymbol U_0^\textsf{H} \boldsymbol \Lambda \boldsymbol
U_0 \bHmH ) \nonumber \\
&& \quad \quad \quad \quad - \sum_{i=1}^{M} \ln ( 1 \!+\! [
\boldsymbol D( \boldsymbol U_0)]_{(i,i)} \lambda_i) \nonumber \\
&\st&  {\rm tr} \{ \boldsymbol \Lambda \} \leq M, \  \lambda_i \geq
0, \ i=1, \ldots, M, \label{newform1.75}
\end{eqnarray}
where $[\boldsymbol D( \boldsymbol U_0)]_{(i,j)}$ denotes the element of the
matrix $\boldsymbol D( \boldsymbol U_0)$ in $i$th row and $j$th column. All
the terms of the objective function and the constraint functions of
\eqref{newform1.75} are concave with respect to $\boldsymbol
\Lambda$ except for the terms $- \ln (1 + [\boldsymbol D (\boldsymbol U_0
)]_{(i,i)} \lambda_i), \, i=1, \ldots, M$, which are convex, and each of
which depends only on a single variable. These convex constraints can be
handled iteratively in terms of their linear approximation around suitably
selected points. Algorithm~1 shows how the POTDC approach can be used for
addressing \eqref{newform1.75}.

\begin{algorithm}[h]                   
\caption{The iterative POTDC algorithm}
\label{alg1}                           
\begin{algorithmic}                    
\REQUIRE Choose the initialization point $\lambda_{i, {\rm c}}  \in [0,M]$,\ \ $i= \ $ \\
\quad \quad \ \ \  $1,\ldots,M$ such that $\sum_{i=1}^M \lambda_{i,{\rm c}}=M$. \\
\quad \quad \ \ \ Select the termination threshold $\zeta_1$,  \\
\quad \quad \ \ \ and set $i$ equal to 1. \REPEAT
  \STATE    Solve the following convex optimization problem using
 $\lambda_{i, {\rm c}}$, $i=1,\ldots,M$ to obtain $\lambda_{i,
 {\rm opt}}$, $i=1,\ldots,M$
\begin{eqnarray}
&\!\!\!\!\!\!\!\!\!\max\limits_{\boldsymbol \Lambda}&\ln \det
(\bI_{\Nm} \!+\! \bHm \boldsymbol U_0^\textsf{H} \boldsymbol \Lambda
\boldsymbol U_0 \bHmH\! ) \nonumber \\
&& \quad\quad \quad \quad - \sum_{i=1}^{M} \frac{[\boldsymbol D (
\boldsymbol U_0)]_{(i,i)} (\lambda_i - \lambda_{i,{\rm c}})}{1 +
[\boldsymbol D (\boldsymbol U_0)]_{(i,i)} \lambda_{i,{\rm c}}}
\nonumber \\
&\st&  {\rm tr} \{ \boldsymbol \Lambda \} \leq M,  \lambda_i \geq
0, \ i=1, \!\ldots\!, M
\end{eqnarray}

and set \STATE  \quad \quad \ \ \ $\boldsymbol \Lambda_{\rm opt}
\leftarrow \boldsymbol \Lambda_{{\rm opt},{ i}} = {\rm
diag}(\lambda_{1,{\rm opt}}, \ldots, \lambda_{M, {\rm opt}})$, \ \
\STATE \quad \quad \ \ \ \ \ $\lambda_{i,{\rm c}} \leftarrow
\lambda_{i, {\rm opt}}$, \ \ \ $i \leftarrow i+1 $
  \UNTIL{$\ $} the difference between two objective values in
  consecutive iterations
    is less than or equal to the termination threshold $\zeta_1$.
\end{algorithmic}
\end{algorithm}

The POTDC algorithm, as applied to \eqref{newform1.75}, is
guaranteed to convergence to a point which satisfies the
Karush-Kuhn-Tucker (KKT) optimality conditions. Moreover, the value
of the objective function is guaranteed to be non-decreasing over the
iterations of the algorithm \cite{khabbazibasmenj2012sum}.

In the next step, we address \eqref{newform2}
with respect to $\boldsymbol U$, when $\boldsymbol \Lambda$ is fixed to
$\boldsymbol \Lambda_0 = \boldsymbol \Lambda_{\rm opt}$ obtained from Algorithm~1. The
corresponding optimization problem can be written as
\begin{eqnarray}
&\max\limits_{\boldsymbol U}& \ln \frac{\det (\bI_{\Nm} + \bHm
\boldsymbol U^\textsf{H} \boldsymbol \Lambda_0 \boldsymbol U
\bHmH )}{\det (\bI_{\Ne} + \bHe \boldsymbol U^H \boldsymbol
\Lambda_0 \boldsymbol U \bHeH )} \nonumber \\
&\st& \boldsymbol U^H \boldsymbol U = \bI_M. \label{secondstep}
\end{eqnarray}
This is an optimization problem over the complex-valued matrix
$\boldsymbol U$ under the constraint that $\boldsymbol U$ is a unitary
matrix. In order to address this problem, we adopt the steepest
descent algorithm on the Lie group of $M \times M$ unitary matrices
developed in~\cite{Koivunen}. This algorithm is shown to
move towards the optimal point over the iterations. To apply
this method, the gradient of the objective function of
\eqref{secondstep} with respect to the complex-valued matrix
$\boldsymbol U$ is required. This gradient can be easily derived as
\begin{align}
\boldsymbol \nabla_{\boldsymbol U} \ln& \frac{\det (\bI_{\Nm} + \bHm
\boldsymbol U^\textsf{H} \boldsymbol \Lambda_0 \boldsymbol U \bHmH )}
{\det (\bI_{\Ne} + \bHe \boldsymbol U^\textsf{H} \boldsymbol
\Lambda_0 \boldsymbol U \bHeH )} \nonumber \\
=&\; \boldsymbol \Lambda_{0} \boldsymbol U \bHmH (\bI_{\Nm} + \bHm
\boldsymbol U^\textsf{H} \boldsymbol \Lambda_{0} \boldsymbol U
\bHmH)^{-1} \bHm \nonumber \\
&- \boldsymbol \Lambda_{0} \boldsymbol U \bHeH (\bI_{\Ne} + \bHe
\boldsymbol U^\textsf{H} \boldsymbol \Lambda_{0} \boldsymbol U
\bHeH)^{-1} \bHe,
\end{align}
where $\boldsymbol \nabla (\cdot) $ stands for the gradient
operator.

The overall algorithm for addressing the precoding matrix design problem
for MIMO Gaussian wire-tap channels can be then described in terms of
Algorithm 2.
\begin{algorithm}[h]                   
\caption{The algorithm for precoding design for MIMO Gaussian wire-tap
channels}                              
\label{alg2}                           
\begin{algorithmic}                    
\REQUIRE Choose arbitrary $\boldsymbol P_{\rm c} = \boldsymbol U_{\rm c}
\boldsymbol \Lambda_{\rm c} \boldsymbol U_{\rm c}^\textsf{H} \succeq
\boldsymbol 0$ such \\
\quad \quad \ \ \ that ${\rm tr}\{ \boldsymbol P_{\rm c} \} = M$. \\
\quad \quad \ \ \ Choose the termination threshold $\zeta_2$, and \\
\quad \quad \ \ \ set $k$ equal to 1. \REPEAT
  \STATE   i) Optimize \eqref{newform1.75} using
  Algorithm~1 for \\ \quad  $\boldsymbol U_{\rm 0} = \boldsymbol
	U_{\rm c}$ and $\boldsymbol \Lambda_{\rm c}$ as the initialization
	point. \\
  \quad $\boldsymbol \Lambda_{\rm c} \leftarrow \boldsymbol
	\Lambda_{\rm opt}$ \\
  ii) Optimize \eqref{secondstep} for $\boldsymbol \Lambda_0
  = \boldsymbol \Lambda_{\rm c}$ to obtain $\boldsymbol U_{\rm opt}$ \\
  \quad $\boldsymbol U_{\rm c} \leftarrow \boldsymbol U_{\rm opt}$, $k
	\leftarrow k+1$ \\
  \UNTIL{$\ $} the difference between two objective values in consecutive
  iterations is less than or equal to the termination threshold $\zeta_2$.
\end{algorithmic}
\end{algorithm}

It is noteworthy to mention that both steps in Algorithm~2,
corresponding to optimization problems \eqref{newform1.75} and
\eqref{secondstep}, result in a non-decreasing value of the objective
function. A detailed convergence analysis and optimality of the proposed
algorithm is left for the future work.

\section{Numerical Examples}\label{sec:numExamples}
\begin{figure}
\centering
  \psfrag{[14]}[][][1]{\cite{khisti2010secure}}
\includegraphics[width=\linewidth]{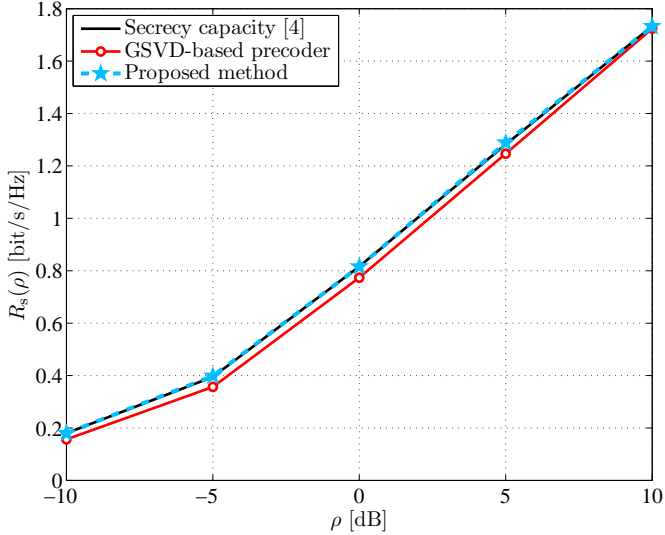}
  \caption{Achievable secrecy rates and the secrecy capacity as
	functions of SNR $\rhom = \rhoe = \rho$ for $M = \Ne = 2$ and $\Nm = 1$.}
  \label{fig:resultCapacityMisomeWiretap}
\end{figure}

In this section, we present the numerical results illustrating our
theoretical findings. The channel matrices $\bHm$ and $\bHe$ are
assumed to be quasi-static flat Rayleigh fading. The displayed results
are averaged over 500 independent channel realizations. The initialization
point for the proposed algorithm, $\boldsymbol P_{\rm c}$, (see Algorithm~2)
is chosen randomly.

For the first scenario, we set $\rhom = \rhoe = \rho$ with $M = \Ne
= 2$ and $\Nm = 1$. The corresponding wire-tap channel is referred to
as multiple-input single-output multi-eavesdropper (MISOME) channel
and its secrecy capacity has been characterized analytically in~\cite{khisti2010misome},
setting a benchmark for the approach proposed in this paper.
Fig.~\ref{fig:resultCapacityMisomeWiretap} compares the secrecy
capacity of the MISOME channel with the achievable secrecy rates
obtained by the GSVD-based beamforming~\cite{fakoorian2012optimal}
and the proposed POTDC method. It can be seen from the figure that
the proposed precoder outperforms the GSVD-based method and
approaches the secrecy capacity derived in~\cite{khisti2010misome}
for the given setup.

\begin{figure}[t!]
\centering
\includegraphics[width=\linewidth]{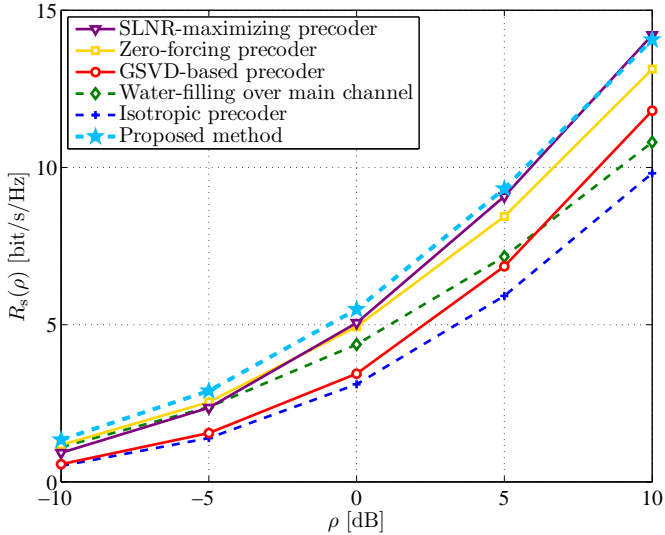}
  \caption{Achievable secrecy rates as
	functions of SNR $\rhom = \rhoe = \rho$ for $M = \Nm = 6$ and $\Ne = 2$.}
  \label{fig:resultMiMimomeWiretap}
    \vspace{-0.5cm}
\end{figure}

For the second scenario, we set $M = \Nm = 6$ and $\Ne = 2$.
Since the capacity is not known for this setting,
we compare our method to the existing alternatives.
Thus, Fig.~\ref{fig:resultMiMimomeWiretap} compares the secrecy rates
achievable by the proposed precoder with those, obtained by
the ZF and SLNR-maximizing  precoders from \cite{wang2012slnr}.
For illustration, we also add conventional water-filling over the main
channel (see, e.g., \cite{cover2012elements}), as well
as an isotropic precoder. It can be seen from the figure that the
latter two are poor strategies in a wire-tap setting. Then, most
importantly, we observe that the proposed approach outperforms
the rest of the strategies.

\section{Conclusions}\label{sec:conclusion}
In this paper, we considered the problem of finding secrecy capacity
of a MIMO wire-tap channel. It has been assumed that the transmitter,
the legitimate receiver, and also the eavesdropper are all equipped
with multiple antennas while the transmitter knows the channel states
from itself to the legitimate receiver and eavesdropper perfectly.
We have developed a novel method for maximizing
the secrecy rate of the aforementioned wire-tap channel. The proposed
algorithm optimizes the eigenvalues and eigenvectors of the
precoding/transmit covariance matrix in alternate manner. Our numerical
results confirm the superiority of the proposed method over the other
state-of-the-art methods.

\bibliographystyle{IEEEtran}

\begin{thebibliography}{1}

\bibitem{wyner1975wire}
 A.~D.~Wyner, ``The wire-tap channel,'' {\it Bell Syst. Tech. J.}, vol.~54, no.~8, pp.~1334-–1387, 1975.

\bibitem{shafiee2009towards}
S.~Shafiee, N.~Liu, and S.~Ulukus, ``Towards the secrecy capacity of
the Gaussian MIMO wire-tap channel: The 2-2-1 channel,'' {\it IEEE
Trans. Inf. Theory}, vol.~55, no.~9, pp.~4033-–4039, Sept.~2009.

\bibitem{liu2009note}
T.~Liu and S.~Shamai, ``A note on the secrecy capacity of the
multiple-antenna wiretap channel,'' {\it IEEE Trans. Inf. Theory},
vol.~55, no.~6, pp.~2547-–2553, Jun.~2009.

\bibitem{khisti2010misome}
A.~Khisti and G.~Wornell, ``Secure transmission with multiple
antennas I: The MISOME wiretap channel,'' {\it IEEE Trans. Inf.
Theory}, vol.~56, no.~7, pp.~3088-–3104, Jul.~2010.

\bibitem{oggier2011secrecy}
F.~Oggier and B.~Hassibi, ``The secrecy capacity of the MIMO
wire-tap channel,'' {\it IEEE Trans. Inf. Theory}, vol.~57, no.~8,
pp.~4961-–4972, Aug.~2011.

\bibitem{li2010optimal}
J.~Li and A.~Petropulu, ``Optimal input covariance for achieving
secrecy capacity in Gaussian MIMO wiretap channels,'' in {\it Proc.
IEEE Int. Conf. Acoust. Speech Sig. Process.}, Dallas, Texas, USA,
Mar.~2010, pp.~3362-–3365.

\bibitem{fakoorian2012optimal}
S.~A.~Fakoorian and A.~L.~Swindlehurst, ``Optimal power allocation
for GSVD-based beamforming in the MIMO Gaussian wiretap channel,''
in {\it Proc. IEEE Int. Symp. Inf. Theory}, Cambridge, MA, USA,
Jul.~2012, pp.~2321-–2325.

\bibitem{wang2012slnr}
K.~Wang, X.~Wang, and X.~Zhang, ``SLNR-based transmit beamforming
for MIMO wiretap channel,'' {\it Wireless Pers. Commun.},
pp.~1–-13, 2012.

\bibitem{khabbazibasmenj2012sum}
 A.~Khabbazibasmenj, F.~Roemer, S.~A.~Vorobyov, and M.~Haardt,
``Sum-rate maximization in two-way AF MIMO relaying: Polynomial time
solutions to a class of DC programming problems,'' {\it IEEE Trans.
Sig. Process.}, vol.~60, no.~10, pp.~5478-–5493, Oct.~2012.

\bibitem{akritas1996various}
A.~G.~Akritas, E.~K.~Akritas, and G.~I.~Malaschonok, ``Various
proofs of {Sylvester's} (determinant) identity,'' {\it Mathematics and
Computers in Simulation}, vol.~42, no.~4, pp.~585–-593, 1996.

\bibitem{Koivunen}
T.~E.~Abrudan, J.~Eriksson, and V.~Koivunen, ``Steepest descent
algorithms for optimization under unitary matrix constraint,'' {\it
IEEE Trans. Sig. Process.}, vol.~56, no.~3, pp.~1134-–1147,
Mar.~2008.

\bibitem{horn2012matrix}
R.~A.~Horn and C.~R.~Johnson, {\it Matrix Analysis}. New York:
Cambridge University Press, 1988.

\bibitem{cover2012elements}
T.~M.~Cover and J.~A.~Thomas, {\it Elements of Information Theory}.
New York: Wiley, 1991.
\end{thebibliography}

\end{document}